\documentclass[vecphys]{svmult}     
\usepackage{graphicx}        
\usepackage{epsfig}          
\usepackage{bm}

\def\chap06a-root-fig{./}

\begin{document}

\title*{Vortices in Bose-Einstein Condensates: Theory}
\author{
N. G. Parker\inst{1} \and B. Jackson\inst{2} \and A. M.
Martin\inst{1} \and C. S. Adams\inst{3}}

\institute{ School of Physics, University of Melbourne, Parkville,
Victoria 3010, Australia.
\texttt{ngparker@ph.unimelb.edu.au,amm@ph.unimelb.edu.au} \and
School of Mathematics and Statistics, Newcastle University,
Newcastle upon Tyne, NE1 7RU, United Kingdom.
\texttt{brian.jackson@newcastle.ac.uk} \and Department of Physics,
Durham University, South Road, Durham, DH1 3LE, United Kingdom.
\texttt{c.s.adams@durham.ac.uk}}

\maketitle

\section{Quantized vortices}
Vortices are pervasive in nature, representing the breakdown of
laminar fluid flow and hence playing a key role in turbulence. The
fluid rotation associated with a vortex can be parameterized by the
circulation $\Gamma=\oint {\rm d}{\bf r}\cdot{\bf v}({\bf r})$ about
the vortex, where ${\bf v}({\bf r})$ is the fluid velocity field.
While classical vortices can take any value of circulation,
superfluids are irrotational, and any rotation or angular momentum is
constrained to occur through vortices with quantized circulation.
Quantized vortices also play a key role in the dissipation of
transport in superfluids.  In BECs quantized vortices have been
observed in several forms, including single vortices
\cite{matthews:prl1999,anderson:prl2000}, vortex lattices
\cite{madison:prl2000,aboshaeer:science2001,hodby:prl2002,raman:prl2001}
(see also Chap.~VII),
and vortex pairs and rings
\cite{anderson:prl2001,dutton:science2001,inouye:prl2001}. The
recent observation of quantized vortices in a fermionic gas was
taken as a clear signature of the underlying condensation and
superfluidity of fermion pairs \cite{zwierlein}. In addition to
BECs, quantized vortices also occur in superfluid Helium
\cite{donnelly,barenghi}, nonlinear optics, and type-II
superconductors \cite{tilley}.

\subsection{Theoretical Framework}

\subsubsection{Quantization of circulation}
Quantized vortices represent phase defects in the superfluid
topology of the system. Under the Madelung transformation, the
macroscopic condensate `wavefunction' $\psi({\bf r},t)$ can be
expressed in terms of a fluid density $n({\bf r},t)$ and a
macroscopic phase $S({\bf r},t)$ via $\psi({\bf r})=\sqrt{n({\bf
r},t)} \exp[iS({\bf r},t)]$. In order that the wavefunction remains
single-valued, the change in phase around any closed contour $C$ must be an
integer multiple of $2\pi$,
\begin{equation}
\int_{\rm C} \nabla S\cdot d{\bf l}=2\pi q,
\end{equation}
where $q$ is an integer.  The gradient of the phase $S$ defines the
superfluid velocity via ${\bf v}({\bf r},t)=(\hbar/m){\bf \nabla}
S({\bf r},t)$. This implies that the circulation about the contour
$C$ is given by,
\begin{equation}
\Gamma=\int_{\rm C} {\bf v}\cdot d{\bf l}=q \left(\frac{h}{m}\right).
\end{equation}
In other words, the circulation of fluid is quantized in units of
$(h/m)$. The circulating fluid velocity about a vortex is given by
${\bf v}(r,\theta)=q\hbar/(mr) \hat{\bm{\theta}}$, where $r$ is the
radius from the core and $\hat{\bm{\theta}}$ is the azimuthal unit
vector.

\subsubsection{Theoretical model}

The Gross-Pitaevskii equation (GPE) provides an excellent
description of BECs at the mean-field level in the limit of
ultra-cold temperature \cite{dalfovo:rmp1999}. It supports quantized
vortices, and has been shown to give a good description of the
static properties and dynamics of vortices
\cite{dalfovo:rmp1999,fetter:jp2001}. Dilute BECs require a
confining potential, formed by magnetic or optical fields,
which typically varies quadratically with position. 
We will assume an
axially-symmetric harmonic trap of the form
$V=\frac{1}{2}m(\omega_r^2 r^2+\omega_z^2 z^2)$, where $\omega_r$
and $\omega_z$ are the radial and axial trap frequencies
respectively.  Excitation spectra of BEC states can be obtained
using the Bogoliubov equations, and specify the stability of
stationary solutions of the GPE. For example, the presence of the
so-called anomalous modes of a vortex in a trapped BEC are
indicative of their thermodynamic instability.  The GPE can also
give a qualitative, and sometimes quantitative, understanding of
vortices in superfluid Helium \cite{donnelly,barenghi}.

Although this Chapter deals primarily with vortices in
repulsively-in\-te\-rac\-ting BECs, vortices in
attractively-interacting BECs have also received theoretical
interest. The presence of a vortex in a trapped BEC with attractive
interactions is less energetically favorable than for repulsive
interactions \cite{dalfovo:pra96}.  Indeed, a harmonically-confined
attractive BEC with angular momentum is expected to exhibit a
center-of-mass motion rather than a vortex \cite{wilkin}. The use of
anharmonic confinement can however support metastable vortices, as
well as regimes of center-of-mass motion and instability
\cite{saito,lundh:prl2004,kavoulakis}.

Various approximations have been made to incorporate thermal effects
into the GPE to describe vortices at finite temperature (see also 
Chap.~XI). The Popov
approximation self-consistently couples the condensate to a normal
gas component using the Bogoliubov-de-Gennes formalism
\cite{virtanen:prl2001} (cf.~Chap.~I Sec.~5.2). Other approaches
involve the addition of thermal/quantum noise to the system, such as
the stochastic GPE method \cite{gardiner,penckwitt,duine:pra2004}
and the classical field/truncated Wigner methods
\cite{steele,davis:pra2002,lobo,simula}. Thermal effects can also be
simulated by adding a phenomenological dissipation term to the GPE
\cite{tsubota}.

\subsubsection{Basic properties of vortices}

In a homogeneous system, a quantized vortex has the 2D form,
\begin{eqnarray}
\psi(r,\theta)=\sqrt{n_{\rm v}(r)}\exp(iq\theta).
\label{vortex-wave}
\end{eqnarray}
The vortex density profile $n_{\rm v}(r)$ has no analytic solution,
although approximate solutions exist \cite{pethick}.  Vortex
solutions can be obtained numerically by propagating the GPE in
imaginary time ($t \rightarrow -it$) \cite{minguzzi04}, whereby the
GPE converges to the lowest energy state of the system (providing it
is stable). By enforcing the phase distribution of
Eq.~(\ref{vortex-wave}), a vortex solution is generated. Figure 1
shows the solution for a $q=1$ vortex at the center of a
harmonically-confined BEC. The vortex consists of a node of zero
density with a width characterized by the condensate healing length
$\xi=\hbar/\sqrt{m n_0 g}$, where $g=4\pi \hbar^2 a /m$ (with $a$
the s-wave scattering length) and $n_0$ is the peak density in the
absence of the vortex.  For typical BEC parameters
\cite{madison:prl2000}, $\xi\sim 0.2~\mu m$.  For a $q=1$ vortex at
the center of an axially-symmetric potential, each particle carries
$\hbar$ of angular momentum. However, if the vortex is off-center,
the angular momentum per particle becomes a function of position
\cite{fetter:jp2001}.

\subsection{Vortex structures}

Increasing the vortex charge widens the core due to centrifugal
effects. In harmonically-confined condensates a multiply-quantized
vortex with $q>1$ is energetically unfavorable compared to a
configuration of singly-charged vortices
\cite{butts:nature99,lundh:pra2002}. Hence, a rotating BEC generally
contains an array of singly-charged vortices in the form of a
triangular Abrikosov lattice
\cite{madison:prl2000,aboshaeer:science2001,hodby:prl2002,raman:prl2001,haljan:prl2001}
(see also Chap.~VII), similar to those found in rotating superfluid
helium \cite{donnelly}. A $q>1$ vortex can decay by splitting into
singly-quantized vortices via a dynamical instability
\cite{mottonen:pra2003,shin:prl2004}, but is stable for some
interaction strengths \cite{pu:pra1999}. Multiply-charged vortices
are also predicted to be stabilized by a suitable localized pinning
potential \cite{simula:pra2002} or the addition of quartic
confinement \cite{lundh:pra2002}.

Two-dimensional vortex-antivortex pairs (i.e.\ two vortices with
equal but opposite circulation) and 3D vortex rings arise in the
dissipation of superflow, and represent solutions to the homogeneous
GPE in the moving frame \cite{jones:jpa1982,jones:jpa1986}, with
their motion being self-induced by the velocity field of the vortex
lines. When the vortex lines are so close that they begin to
overlap, these states are no longer stable and evolves into a
rarefaction pulse \cite{jones:jpa1982}.

Having more than one spin component in the BECs (cf.~Chap.~IX)
provides an additional topology to vortex structures.  Coreless
vortices and vortex `molecules' in coupled two-component BECs have
been probed experimentally \cite{leanhardt:prl2003} and
theoretically \cite{kasamatsu:prl2004}.   More exotic vortex
structures such as skyrmion excitations \cite{ruostekoski} and
half-quantum vortex rings \cite{ruostekoski:prl2003} have also been
proposed.

\section{Nucleation of vortices}

Vortices can be generated by rotation, a moving obstacle, or phase imprinting
methods. Below we discuss each method in turn.

\subsection{Rotation}

As discussed in the previous section, a BEC can only rotate through
the existence of quantized vortex lines. Vortex nucleation occurs
only when the rotation frequency $\Omega$ of the container exceeds a
critical value $\Omega_c$
\cite{fetter:jp2001,butts:nature99,nozieres}. Consider a condensate
in an axially-symmetric trap which is rotating about the {\it
z}-axis at frequency $\Omega$. In the Thomas-Fermi limit, the
presence of a vortex becomes energetically favorable when $\Omega$
exceeds a critical value given by \cite{Lundh97},
\begin{equation}
\Omega_c=\frac{5}{2}\frac{\hbar}{mR^2} \ln \frac{0.67 R}{\xi}.
\label{Omega_c}
\end{equation}
This is derived by integrating the kinetic energy density $m n(r)
v(r)^2/2$ of the vortex velocity field in the radial plane. The
lower and upper limits of the integration are set by the healing
length $\xi$ and the BEC Thomas-Fermi radius $R$, respectively. Note
that $\Omega_c<\omega_r$ for repulsive interactions, while
$\Omega_c>\omega_r$ for attractive interactions
\cite{dalfovo:pra96}. In a non-rotating BEC the presence of a vortex
raises the energy of the system, indicating thermodynamic
instability \cite{rokhsar:prl1997}.

In experiments, vortices are formed only when the trap is rotated at
a much higher frequency than $\Omega_c$
\cite{madison:prl2000,aboshaeer:science2001,hodby:prl2002},
demonstrating that the energetic criterion is a necessary, but not
sufficient, condition for vortex nucleation. There must also be a
dynamic route for vorticity to be introduced into the condensate,
and hence Eq.\ (\ref{Omega_c}) provides only a lower bound for the
critical frequency.

The nucleation of vortices in rotating trapped BECs appears to be
linked to instabilities of collective excitations. Numerical
simulations based on the GPE have shown that once the amplitude of
these excitations become sufficiently large, vortices are nucleated
that subsequently penetrate the high-density bulk of the condensate
\cite{penckwitt,lobo,tsubota,lundh:pra2003,parker:lattice}.

One way to induce instability is to resonantly excite a surface mode
by adding a rotating deformation to the trap potential. In the limit
of small perturbations, this resonance occurs close to a rotation
frequency $\Omega_r = \omega_\ell /\ell$, where $\omega_\ell$ is the
frequency of a surface mode with multipolarity $\ell$. In the
Thomas-Fermi limit, the surface modes satisfy $\omega_\ell
=\sqrt{\ell}\omega_r$ \cite{stringari96}, so $\Omega_r =
\omega_r/\sqrt{\ell}$. For example, an elliptically-deformed trap,
which excites the $\ell=2$ quadrupole mode, would nucleate vortices
when rotated at $\Omega_r \approx \omega_r/\sqrt{2}$. This value has
been confirmed in both experiments
\cite{madison:prl2000,aboshaeer:science2001,hodby:prl2002} and
numerical simulations
\cite{penckwitt,lobo,tsubota,lundh:pra2003,parker:lattice}. Higher
multipolarities were resonantly excited in the experiment of
Ref.~\cite{raman:prl2001}, finding vortex formation at frequencies
close to the expected values, $\Omega = \omega_r/\sqrt{\ell}$, and
lending further support to this picture.

A similar route to vortex nucleation is revealed by considering
stationary states of the BEC in a rotating elliptical trap, which
can be obtained in the Thomas-Fermi limit by solving hydrodynamic
equations \cite{recati01}. At low rotation rates only one solution
is found; however at higher rotations ($\Omega > \omega_r/\sqrt{2}$)
a bifurcation occurs and up to three solutions are present. Above
the bifurcation point one or more of the solutions become
dynamically unstable \cite{Sinha01}, leading to vortex formation
\cite{Parker06}. Madison {\it et al.\ }\cite{madison01} followed
these stationary states experimentally by adiabatically introducing
trap ellipticity and rotation, and observed vortex nucleation in the
expected region.

Surface mode instabilities can also be induced at finite temperature
by the presence of a rotating noncondensed ``thermal'' cloud. Such
instabilities occur when the thermal cloud rotation rate satisfies
$\Omega > \omega_{\ell} /\ell$ \cite{williams02}. Since all modes can
potentially be excited in this way, the criterion for instability
and hence vortex nucleation becomes $\Omega_c > {\rm min}
(\omega_{\ell}/\ell)$, analogous to the Landau criterion. Note that such a
minimum exists at $\Omega_c >0$ since the Thomas-Fermi result
$\omega_\ell =\sqrt{\ell}\omega_r$ becomes less accurate for high $\ell$
\cite{dalfovo01}. This mechanism may have been important in the
experiment of Haljan {\it et al.\ }\cite{haljan:prl2001}, where a
vortex lattice was formed by cooling a rotating thermal cloud to
below $T_c$.

\subsection{Nucleation by a moving object}

Vortices can also be nucleated in BECs by a moving localized
potential. This problem was originally studied using the GPE for 2D
uniform condensate flow around a circular hard-walled potential
\cite{frisch92,winiecki99}, with vortex-antivortex pairs being
nucleated when the flow velocity exceeded a critical value.

In trapped BECs a similar situation can be realized using the
optical dipole force from a laser, giving rise to a localized
repulsive Gaussian potential. Under linear motion of such a
potential, numerical simulations revealed vortex pair formation when
the potential is moved at a velocity above a critical value \cite{jackson98}.
The experiments of \cite{raman99,onofrio00} oscillated a repulsive laser
beam in an elongated condensate. Although vortices were not observed
directly, the measurement of condensate heating and drag above a
critical velocity was consistent with the nucleation of vortices
\cite{jackson:pra2000a}.

An alternative approach is to move the laser beam potential in a
circular path around the trap center \cite{caradoc99}. By ``stirring'' the
condensate in this way one or more vortices can be created. This
technique was used in the experiment of Ref.~\cite{raman:prl2001},
where vortices were generated even at low stirring frequencies.

\subsection{Other mechanisms and structures}

A variety of other schemes for vortex creation have been suggested.
One of the most important is that by Williams and Holland
\cite{williams99}, who proposed a combination of rotation and
coupling between two hyperfine levels to create a two-component
condensate, one of which is in a vortex state. The non-vortex
component can then either be retained or removed with a resonant
laser pulse. This scheme was used by the first experiment to obtain
vortices in BEC \cite{matthews:prl1999}. A related method, using
topological phase imprinting, has been used to experimentally
generate multiply-quantized vortices \cite{leanhardt:prl2002}.

Apart from the vortex lines considered so far, vortex rings have also
been the subject of interest. Rings are the decay
product of dynamically unstable dark solitary waves in 3D geometries
\cite{anderson:prl2001,dutton:science2001,ginsberg:prl2005,komineas:pra2003}.
Vortex rings also form
in the quantum reflection of BECs from surface potentials
\cite{scott:prl2005}, the unstable motion of BECs through an optical
lattice \cite{scott:pra2004}, the dragging of a 3D object through a
BEC \cite{jackson:pra1999}, and the collapse of ultrasound bubbles
in BECs \cite{berloff:prl2004}. The controlled generation of vortex
rings \cite{ruostekoski:pra2005} and multiple/bound vortex ring
structures \cite{crasovan} have been analyzed theoretically.

A finite temperature state of a quasi-2D BEC, characterized by the
thermal activation of vortex-antivortex pairs, has been simulated
using classical field simulations \cite{simula:prl2006}.  This
effect is thought to be linked to the
Berezinskii-Kosterlitz-Thouless phase transition of 2D superfluids,
recently observed experimentally in ultracold gases \cite{hadzibabic06}.
Similar simulations in a 3D system have also demonstrated the
thermal creation of vortices \cite{davis02,berloff02}.

\section{Dynamics of vortices}

The study of vortex dynamics has long been an important topic in
both classical \cite{lamb} and quantum \cite{barenghi}
hydrodynamics. Helmholtz's theorem for uniform, inviscid fluids,
which is also applicable to quantized vortices in superfluids near
zero temperature, states that the vortex will follow the motion of
the background fluid. So, for example, in a superfluid with uniform
flow velocity ${\bf v}_s$, a single straight vortex line will move
with velocity ${\bf v}_L$, such that it is stationary in the frame
of the superfluid.

Vortices similarly follow the ``background flow'' originating from
circulating fluid around a vortex core. Hence vortex motion can be
induced by the presence of other vortices, or by other parts of the
same vortex line when it is curved. Most generally, the superfluid
velocity ${\bf v}_i$ due to vortices at a particular point ${\bf r}$
is given by the {\it Biot-Savart} law \cite{barenghi}, in analogy
with the similar equation in electromagnetism,
\begin{equation}
 {\bf v}_i = \frac{\Gamma}{4 \pi} \int \frac{({\bf s}-{\bf r}) \times d{\bf s}}
 {|{\bf s}-{\bf r}|^3};
\label{biot-savart}
\end{equation}
where ${\bf s} (\zeta,t)$ is a curve representing the vortex line
with $\zeta$ the arc length. Equation (\ref{biot-savart}) suffers
from a divergence at ${\bf r}={\bf s}$, so in calculations of vortex
dynamics this must be treated carefully \cite{tsubota00}. Equation
(\ref{biot-savart}) also assumes that the vortex core size is small
compared to the distance between vortices. In particular, it breaks
down when vortices cross during collisions, where reconnection
events can occur. These reconnections can either be included
manually \cite{schwarz85}, or by solving the full GPE
\cite{koplik93}. The latter method also has the advantage of
including sound emission due to vortex motion or reconnections
\cite{leadbeater:prl2001,leadbeater:pra2003}.

In a system with multiple vortices, motion of one vortex is induced
by the circulating fluid flow around other vortices, and vice-versa
\cite{donnelly}. This means that, for example, a pair of vortices of
equal but opposite charge will move linearly and parallel to each
other with a velocity inversely proportional to the distance between
them. Two or more vortices of equal charge, meanwhile, will rotate
around each other, giving rise to a rotating vortex lattice as will
be discussed in Chap.~VII. When a vortex line is curved,
circulating fluid from one part of the line can induce motion in
another. This effect can give rise to helical waves on the vortex,
known as Kelvin modes \cite{kelvin}. It also has interesting
consequences for a vortex ring, which will travel
in a direction perpendicular to the plane of the ring, with a
self-induced velocity that decreases with increasing radius.
Classically, this is most familiar in the motion of smoke rings,
though similar behavior has also been observed in superfluid helium
\cite{rayfield64}.

This simple picture is complicated in the presence of density
inhomogeneities or confining walls. In a harmonically-trapped BEC
the density is a function of position, and therefore the energy, $E$, of a
vortex will also depend on its position within the condensate.
To simplify matters, let us consider a quasi-2D situation, where the
condensate is pancake-shaped and the vortex line is straight. In
this case, the energy of the vortex depends on its displacement
${\bf r}$ from the condensate center \cite{svidzinsky:prl2000}, and
a displaced vortex feels a force proportional to $\nabla E$. This is
equivalent to a Magnus force on the vortex
\cite{jackson:pra2000b,lundh00,mcgee01} and to compensate the vortex
moves in a direction perpendicular to the force, leading it to
precess around the center of the condensate along a line of
constant energy. This precession of a single vortex has been
observed experimentally \cite{anderson:prl2000}, with a frequency in
agreement with theoretical predictions. In more 3D
situations, such as spherical or cigar-shaped condensates, the
vortex can bend
\cite{garcia:pra2001a,garcia:pra2001b,aftalion:pra2001,rosenbusch:prl2002}
leading to more complicated motion \cite{fetter:jp2001}. Kelvin
modes \cite{bretin:prl2003,fetter:pra2004} and vortex ring dynamics
\cite{jackson:pra2000b} are also modified by the density
inhomogeneity in the trap.

In the presence of a hard-wall potential, a new constraint is imposed such
that the fluid velocity normal to the wall must be zero,
${\bf v}_s \cdot \hat{\bf n}=0$. The resulting problem of vortex motion is
usually solved mathematically \cite{lamb} by invoking an ``image vortex''
on the other side of the wall (i.e.\ in the region where there is no
fluid present), at a position such that its normal flow cancels that of
the real vortex at the barrier. The motion of the real vortex is then
simply equal to the induced velocity from the image vortex circulation.

\section{Stability of vortices}

\subsection{Thermal instabilities}

At finite temperatures the above discussion is modified by the
thermal occupation of excited modes of the system, which gives rise to a
noncondensed normal fluid in addition to the superfluid. 
A vortex core moving relative to the normal fluid scatters thermal
excitations, and will therefore feel a frictional force leading
to dissipation. This mutual friction force can be written as \cite{donnelly},
\begin{equation}
 {\bf f}_D = - n_s \Gamma \{ \alpha {\bf s}' \times
 [\, {\bf s}' \times ({\bf v}_n - {\bf v}_L)]
 + \alpha' {\bf s}' \times ({\bf v}_n - {\bf v}_L) \},
\label{eq:mutual}
\end{equation}
where $n_s$ is the background superfluid density, ${\bf s}'$ is the
derivative of ${\bf s}$ with respect to arc length $\zeta$, $\alpha$
and $\alpha'$ are temperature dependent parameters, while ${\bf
v}_L$ and ${\bf v}_n$ are the velocities of the vortex line and
normal fluid respectively. The mutual friction therefore has two
components perpendicular to the relative velocity ${\bf v}_n-{\bf
v}_L$.

To consider an example discussed in the last section, an off-center
vortex in a trapped BEC at zero temperature will precess such that
its energy remains constant. In the presence of a non-condensed
component, however, dissipation will lead to a loss of energy. Since
the vortex is topological it cannot simply vanish, so this lost
energy is manifested as a radial drift of the vortex towards lower
densities. In Eq.~(\ref{eq:mutual}) the $\alpha$ term  is
responsible for this radial motion, while $\alpha'$ changes the
precession frequency.  The vortex disappears at the edge of the
condensate, where it is thought to decay into elementary excitations
\cite{fedichev:pra1999vor}. Calculations based upon the stochastic GPE 
have shown that thermal fluctuations lead to an uncertainty in the
position of the vortex, such that even a central vortex will
experience thermal dissipation and have a finite lifetime
\cite{duine:pra2004}. This thermodynamic lifetime is predicted to be of 
the order of seconds
\cite{fedichev:pra1999vor}, which is consistent with experiments
\cite{matthews:prl1999,madison:prl2000,rosenbusch:prl2002}.

\subsection{Hydrodynamic instabilities}
Experiments indicate that the crystallization of vortex lattices is
temperature-independent \cite{hodby:prl2002,aboshaeer:prl2002}.
Similarly, vortex tangles in turbulent states of superfluid Helium
have been observed to decay at ultracold temperature, where thermal
dissipation is virtually nonexistent \cite{davis:pb2000}. These
results highlight the occurrence of zero temperature dissipation
mechanisms, as listed below.

\subsubsection{Instability to acceleration}
The topology of a 2D homogeneous superfluid can be mapped on to a
(2+1)D electrodynamic system, with vortices and phonons playing the
role of charges and photons respectively \cite{arovas}. Just as an
accelerating electron radiates according to the Larmor acceleration 
squared law, a superfluid vortex
is inherently unstable to acceleration and radiates sound waves.

\begin{figure}[h!]
\center
\includegraphics[width=11cm,clip=true]{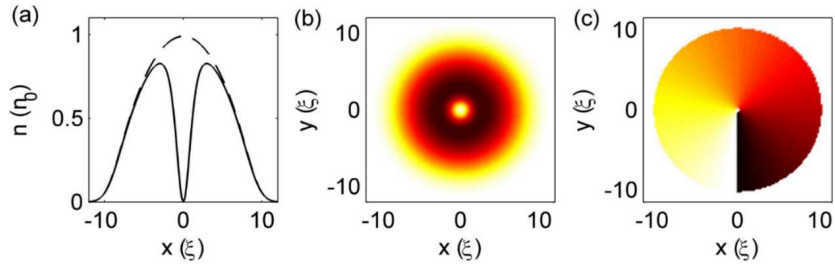}
\caption{Profile of a singly-quantized ($q=1$) vortex at the center of
a harmonically-confined BEC: (a) condensate density along the $y=0$ axis
(solid line) and the corresponding density profile in the absence of the
vortex (dashed line). (b) 2D density and (c) phase profile of the vortex
state. These profiles are calculated
 numerically by propagating the 2D GPE in imaginary time subject to an
 azimuthal $2\pi$ phase variation around the trap center.}
\label{vortex_soliton_profile}
\end{figure}

\begin{figure}[h!]
\centering
\includegraphics[width=10cm]{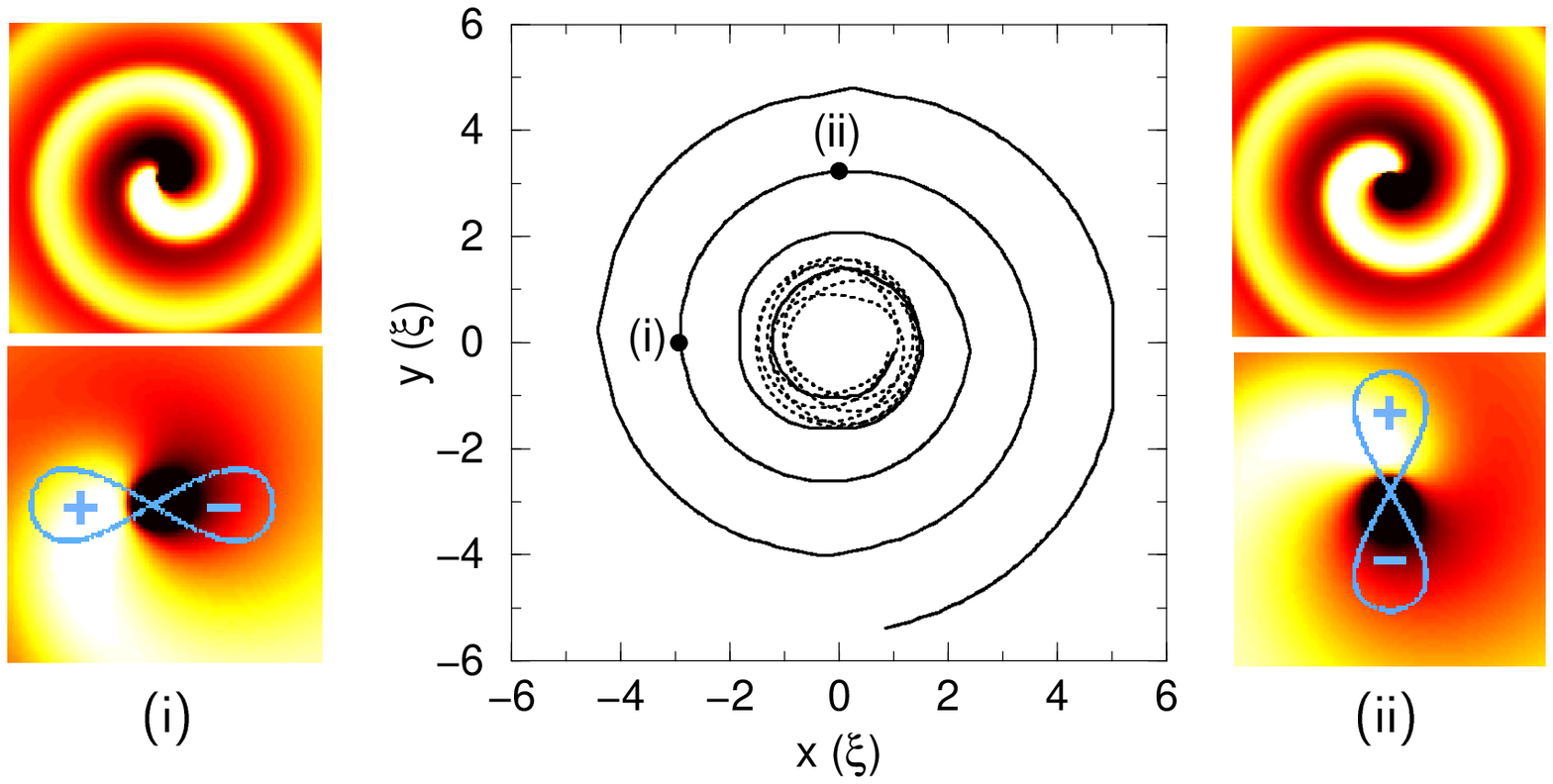}
\caption{Vortex path in the dimple trap geometry of
Eq.~(\ref{eqn:dimple}) with $\omega_{\rm d}=0.28 (c/\xi)$. Deep $V_0
=10\mu$ dimple (dotted line): mean radius is constant, but modulated
by the sound field. Shallow $V_0 =0.6\mu$ dimple and homogeneous
outer region $\omega_r=0$ (dotted line): vortex spirals outwards.
Outer plots: Sound excitations (with amplitude $\sim 0.01n_0$)
radiated in the $V_0=0.6\mu$ system at times indicated. Top:
Far-field distribution $[-90,90]\xi \times[-90,90]\xi$. Bottom:
Near-field distribution $[-25,25]\xi \times[-25,25]\xi$, with an
illustration of the dipolar radiation pattern. Copyright (2004) by
the American Physical Society \cite{parker:prl2004}.}
\label{vortex_spiral}
\end{figure}

Vortex acceleration can be induced by the presence of an
inhomogeneous background density, such as in a trapped BEC. Sound
emission from a vortex in a BEC can be probed by considering a trap
of the form \cite{parker:prl2004},
\begin{equation}
V_{\rm ext}=V_0\left[1-\exp\left(- \frac{m\omega_{\rm d}^2
r^2}{2V_0} \right)\right]+ \frac{1}{2}m\omega_r^2 r^2.
\label{eqn:dimple}
\end{equation}
This consists of a gaussian dimple trap with depth $V_0$ and
harmonic frequency component $\omega_{\rm d}$, embedded in an
ambient harmonic trap of frequency $\omega_r$. A 2D description is
sufficient to describe this effect.  This set-up can be realized
with a quasi-2D BEC by focussing a far-off-resonant red-detuned
laser beam in the center of a magnetic trap. The vortex is initially
confined in the inner region, where it precesses due to the
inhomogeneous density.
Since sound excitations have an energy
of the order of the chemical potential $\mu$, the depth of the
dimple relative to $\mu$ leads to two distinct regimes of
vortex-sound interactions.

$V_0\gg \mu$: The vortex effectively sees an infinite harmonic
trap - it precesses and radiates sound but there is no net decay
due to complete sound reabsorption. However, a collective mode of
the background fluid is excited, inducing slight modulations in the
vortex path (dotted line in Fig~\ref{vortex_spiral}).

$V_0<\mu$: Sound waves are radiated by the precessing vortex.
Assuming $\omega_r=0$, the sound waves propagate to infinity without
reinteracting with the vortex. The ensuing decay causes the vortex
to drift to lower densities, resulting in a spiral motion (solid
line in Fig.~\ref{vortex_spiral}), similar to the effect of thermal
dissipation. The sound waves are emitted in a dipolar radiation
pattern, perpendicularly to the instantaneous direction of motion
(subplots in Fig.~\ref{vortex_spiral}), with a typical amplitude of
order $0.01 n_0$ and wavelength $\lambda \sim 2\pi c/\omega_{\rm V}$
\cite{fetter:jp2001}, where $c$ is the speed of sound and $\omega_V$
is the vortex precession frequency. The power radiated from a vortex
can be expressed in the form
\cite{parker:prl2004,vinen:prb2000,lundh:pra2000},
\begin{eqnarray}
P=\beta  m N \left(\frac{a^2}{\omega_{\rm V}}\right),
\label{eqn:vortex_power}
\end{eqnarray}
where $a$ is the vortex acceleration, $N$ is the total number of
atoms, and $\beta$ is a dimensionless coefficient.  Using classical
hydrodynamics \cite{vinen:prb2000} and by mapping the superfluid
hydrodynamic equations onto Maxwell's electrodynamic equations
\cite{lundh:pra2000}, it has been predicted that $\beta=\pi^2/2$
under the assumptions of a homogeneous 2D fluid, a point vortex, and
perfect circular motion. Full numerical simulations of the GPE based
on a realistic experimental scenario have derived a coefficient of
$\beta \sim 6.3 \pm 0.9$ (one standard deviation), with the
variation due to a weak dependence on the geometry of the system
\cite{parker:prl2004}.

When $\omega_r \neq 0$, the sound eventually reinteracts with the
vortex, slowing but not preventing the vortex decay. By varying $V_0$ it is
possible to
control vortex decay, and in suitably engineered traps this decay
mechanism is expected to dominate over thermal dissipation
\cite{parker:prl2004}.

Vortex acceleration (and sound emission) can also be induced by the
presence of other vortices.  A co-rotating pair of two vortices of
equal charge has been shown to decay continuously via quadrupolar
sound emission, both analytically \cite{pismen} and numerically
\cite{barenghi:jltp2004}. Three-body vortex interactions in the form
of a vortex-antivortex pair incident on a single vortex have also
been simulated numerically, with the interaction inducing
acceleration in the vortices with an associated emission of sound
waves \cite{barenghi:jltp2004}.

Simulations of vortex lattice formation in a rotating elliptical
trap show that vortices are initially nucleated in a turbulent
disordered state, before relaxing into an ordered lattice
\cite{parker:lattice}.  This relaxation process is associated with
an exchange of energy from the sound field to the vortices due to
these vortex-sound interactions. This agrees with the experimental
observation that vortex lattice formation is insensitive to
temperature \cite{hodby:prl2002,aboshaeer:prl2002}.

\subsubsection{Kelvin wave radiation and vortex reconnections}
In 3D a Kelvin wave excitation will induce acceleration in the
elements of the vortex line, and therefore local sound emission.
Indeed, simulations of the GPE in 3D have shown that Kelvin waves
excitations on a vortex ring lead to a decrease in the ring size,
indicating the underlying radiation process
\cite{leadbeater:pra2003}.  Kelvin wave excitations can be generated
from a vortex line reconnection
\cite{leadbeater:prl2001,leadbeater:pra2003} and the interaction of
a vortex with a rarefaction pulse \cite{berloff:pra2004}.

Vortex lines which cross each other can undergo dislocations and
reconnections \cite{caradoc}, which induce
a considerable burst of sound emission \cite{leadbeater:prl2001}.
Although they have yet to be probed experimentally in BECs, vortex
reconnections are hence thought to play a key role in the dissipation of
vortex tangles in Helium II at ultra-low temperatures
\cite{donnelly}.

\section{Dipolar BECs}

A BEC has recently been formed of chromium atoms
\cite{Griesmaier05}, which feature a large dipole moment. This
opens the door to studying of the effect of long-range dipolar
interactions in BECs.

\subsection{The Modified Gross-Pitaevskii Equation \label{MGPE}}

The interaction potential $U_{dd}({\bf r})$ between two dipoles
separated by $\bf{r}$, and aligned by an external field along the
unit vector $\hat{\bf{e}}$ is given by,
\begin{eqnarray}
U_{dd}({\bf r})=\frac{C_{dd}}{4 \pi} \hat{e}_i\hat{e}_j
\frac{\left(\delta_{ij}-3\hat{r}_i \hat{r}_j\right)}{r^3}.
\label{eqn:U_dd_Dipolar}
\end{eqnarray}
For low energy scattering of two atoms with dipoles induced by a
static electric field ${\bf E}=E \hat{\bf{e}}$, the coupling
constant $C_{dd}=E^2 \alpha^2/\epsilon_0$ \cite{Marinescu98,Yi00},
where $\alpha$ is the static dipole polarizability of the atoms and
$\epsilon_0$ is the permittivity of free space. Alternatively, if
the atoms have permanent magnetic dipoles, $d_m$, aligned in an
external magnetic field ${\bf B}=B \hat{\bf{e}}$, one has $C_{dd}=
\mu_0 d_m^2$ \cite{Goral00}, where $\mu_0$ is the permeability of
free space. Such dipolar interactions give rise to a mean-field
potential
\begin{equation}
\Phi_{dd} ({\bf r}) = \int d^3 r U_{dd} \left( {\bf r}-{\bf
r}^{\prime} \right) |\psi\left({\bf r}^{\prime} \right)|^2,
\label{eqn:Phi_dd_Dipolar}
\end{equation}
which can be incorporated into the GPE to give,
\begin{equation}\label{GPE_Dipolar}
i \hbar \psi_t = \left[ -\frac{\hbar^2}{2m} \nabla^2 + g|\psi|^2 +
\Phi_{dd} + V\right]\psi.
\end{equation}
For an axially-symmetric quasi-2D geometry ($\omega_z\gg\omega_r$)
rotating about the ${\it z}$-axis, the ground state wavefunction of
a single vortex has been solved numerically \cite{Yi06}. Considering
$10^5$ chromium atoms and $\omega_r=2\pi \, \times 100$Hz,
several solutions were obtained depending on the strength of the
$s$-wave interactions and the alignment of the dipoles relative to
the trap.

For the case of axially-polarized dipoles the most striking results
arise for attractive $s$-wave interactions $g<0$. Here the BEC
density is axially symmetric and oscillates in the vicinity of the
vortex core. Similar density oscillations have been observed in
numerical studies of other non-local interaction potentials,
employed to investigate the interparticle interactions in $^4$He
\cite{Oritz95,Sadd97,Berloff99,Dalfovo92}, with an interpretation
that relates to the roton structure in a superfluid
\cite{Dalfovo92}. For the case of transversely-polarized dipoles,
where the polarizing field is co-rotating with the BEC, and
repulsive $s$-wave interactions ($g>0$), the BEC becomes elongated
along the axis of polarization \cite{Stuhler05} and as a consequence
the vortex core is anisotropic.

\subsection{Vortex Energy \label{Vortex_Energy}}

Assuming a dipolar BEC in the TF limit (cf.~Sec.~5.1 in Chap.~I),
the energetic cost of a vortex, aligned along the axis of
polarization ($z$-axis), has been derived using a variational ansatz
for the vortex core \cite{ODell06}, and thereby the critical
rotation frequency $\Omega_c$ at which the presence of a vortex
becomes energetically favorable has been calculated. For an oblate
trap ($\omega_{r}< \omega_z$), dipolar interactions decrease
$\Omega_c$, while for prolate traps ($\omega_{r}
> \omega_z$) the presence of dipolar interactions
increases $\Omega_c$. A formula resembling Eq. (\ref{Omega_c}) for
the critical frequency of a conventional BEC can be used to explain
these results, with $R$ being the modified TF radius of the dipolar
BEC. Indeed, using the TF radius of a vortex-free dipolar BEC
\cite{ODell04,Eberlein05} and the conventional $\it s$-wave healing
length $\xi$, it was found that Eq. (\ref{Omega_c}) closely matches
the results from the energy cost calculation. Deviations become
significant when the dipolar interactions dominate over {\it s}-wave
interactions. In this regime the $\it s$-wave healing length $\xi$
is no longer the relevant length scale of the system, and the
equivalent dipolar length scale $\xi_d=C_{dd}m/(12 \pi \hbar^2)$
will characterize the vortex core size.

For $g>0$ and in the absence of dipolar interactions, the rotation
frequency at which the vortex-free BEC becomes dynamically unstable,
$\Omega_{dyn}$, is always greater than the critical frequency for
vortex stabilization $\Omega_c$. However in the presence of dipolar
interactions, $\Omega_{dyn}$ can become less than $\Omega_c$,
leading to an intriguing regime in which the dipolar BEC is
dynamically unstable but vortices will not enter
\cite{ODell06,Bijnen06}. As with attractive condensates
\cite{wilkin}, the angular momentum may then be manifested as center
of mass oscillations.

\section{Analogs of Gravitational Physics in BECs}

There is growing interest in pursuing analogs of gravitational
physics in condensed matter systems \cite{Barcelo05}, such as BECs.
The rationale behind such models can be traced back to the work of
Unruh \cite{Unruh81,Unruh95}, who noted the analogy between sound
propagation in an inhomogeneous background flow and field
propagation in curved space-time. This link applies in the TF limit
of BECs where the speed of sound is directly analogous to the speed
of light in the corresponding gravitational system \cite{Barcelo01}.
This has led to proposals for experiments to probe effects such as
Hawking radiation \cite{Hawking74,Hawking75} and superradiance
\cite{Bekenstein98}. For Hawking radiation it is preferable to avoid
the generation of vortices \cite{Barcelo05,Barcelo03}, and as such
will not be discussed here. However, the phenomena of superradiance
in BECs, which can be considered as stimulated Hawking radiation,
relies on the presence of a vortex
\cite{Slatyer05,Basak03a,Basak03b,Federici06}, which is analogous to
a rotating black hole.

Below we outline the derivation of how the propagation of sound in a
BEC can be considered to be analogous to field propagation
\cite{Barcelo05}. From the GPE it is possible to derive the
continuity equation for an irrotational fluid flow with phase
$S({\bf r},t)$ and density $n({\bf r},t)$, and a Hamilton-Jacobi
equation whose gradient leads to the Euler equation. Linearizing
these equations with respect to the background it is found that
\begin{equation}
\label{Linearization1}
\partial_t S'=-\frac{1}{m}\nabla S \cdot \nabla
S'-gn'+\frac{\hbar^2}{4m\sqrt{n}}
\left(\nabla^2\frac{n'}{\sqrt{n}}-\frac{n'}{n}\nabla^2\sqrt{n}\right),
\end{equation}
\begin{equation}
\label{Linearization2}
\partial_t n'=-\frac{1}{m}\nabla \cdot \left(n \nabla
S'\right)-\frac{1}{m}\nabla \cdot \left(n' \nabla S\right),
\end{equation}
where $n'$ and $S'$ are the perturbed values of the density $n$ and
phase $S$ respectively. Neglecting the quantum pressure
$\nabla^2$-terms, the above equations can be rewritten as a
covariant differential equation describing the propagation of phase
oscillations in a BEC. This is directly analogous to the propagation
of a minimally coupled massless scalar field in an effective
Lorentzian geometry which is determined by the background velocity,
density and speed of sound in the BEC. Hence, the propagation of
sound in a BEC can be used as an analogy for the propagation of
electromagnetic fields in the corresponding space-time. Of course
one has to be aware that this direct analogy is only valid in the TF
regime, which breaks down on scales of the order of a healing
length, i.e. the theory is only valid on large length scales, as is
general relativity.

\subsection{Superradiance}
Superradiance in BECs relies on sound waves incident on a vortex
structure and is characterized by the reflected sound energy
exceeding the incident energy.  This has been studied using Eqs.
(\ref{Linearization1}) and (\ref{Linearization2}) for monochromatic
sound waves of frequency $\omega_s$ and angular wave number $q_s$
incident upon a vortex \cite{Slatyer05} and a `draining vortex'' (a
vortex with outcoupling at its center)
\cite{Basak03a,Basak03b,Federici06}.

For the vortex case, a vortex velocity field ${\bf
v}(r,\theta)=(\beta/r)\hat{\bm{\theta}}$ and a density profile
ansatz was assumed.  Superradiance then occurs when $\beta q_s>A
c_\infty$, where $A$ is related to the vortex density ansatz and
$c_\infty$ is the speed of sound at infinity \cite{Slatyer05}.
Interestingly, this condition is frequency independent.

For the case of a draining vortex, an event horizon occurs at a
distance $a$ from the vortex core, where the fluid circulates at
frequency $\Omega$. Assuming a homogeneous density $n$ and a
velocity profile ${\bf v}(r,\theta)=\left(-ca \hat{\bf r}+\Omega a^2
\hat{ {\bm \theta}}\right)/r$ where $c$ is the homogeneous speed of
sound, superradiance occurs when $0 < \omega_s < q_s \Omega$
\cite{Basak03a,Basak03b,Federici06}.

\nopagebreak
The increase in energy of the outgoing sound is due to an extraction
of energy from the vortex and as such it is expected to lead to
slowing of the vortex rotation. However, such models do not
include quantized vortex angular momentum, and as such it is
expected that superradiance will be suppressed \cite{Federici06}.
This raises tantalizing questions, such as whether superradiance can
occur if vorticity is quantized, if such effects can be modeled with
the GPE, and whether the study of quantum effects in condensate
superradiance will shed light on quantum effects in general
relativity.



\end{document}